\documentclass[10pt,twocolumn,letterpaper]{article}

\usepackage{3dv}
\usepackage{times}
\usepackage{epsfig}
\usepackage{graphicx}
\usepackage{amsmath}
\usepackage{amssymb}
\usepackage{float}

\usepackage[pagebackref=true,breaklinks=true,colorlinks,bookmarks=false]{hyperref}

\threedvfinalcopy % *** Uncomment this line for the final submission

\def\threedvPaperID{223} % *** Enter the 3DV Paper ID here

\ifthreedvfinal\fi
\begin{document}

%%%%%%%%% TITLE
\title{Neural Implicit Surfaces for Efficient and Accurate Collisions in Physically Based Simulations}

\author{Hugo Bertiche\\
Universitat de Barcelona\\
Computer Vision Center\\
{\tt\small hugo\_bertiche@hotmail.com}
\and
Meysam Madadi\\
Universitat de Barcelona\\
Computer Vision Center\\

\and
Sergio Escalera\\
Universitat de Barcelona\\
Computer Vision Center\\}

\maketitle
% \thispagestyle{empty}

%%%%%%%%% ABSTRACT
\begin{abstract}
   Current trends in the computer graphics community propose leveraging the massive parallel computational power of GPUs to accelerate physically based simulations. Collision detection and solving is a fundamental part of this process. It is also the most significant bottleneck on physically based simulations and it easily becomes intractable as the number of vertices in the scene increases. \textit{Brute force} approaches carry a quadratic growth in both computational time and memory footprint. While their parallelization is trivial in GPUs, their complexity discourages from using such approaches. Acceleration structures --such as BVH-- are often applied to increase performance, achieving logarithmic computational times for individual point queries. Nonetheless, their memory footprint also grows rapidly and their parallelization in a GPU is problematic due to their branching nature. We propose using implicit surface representations learnt through deep learning for collision handling in physically based simulations. Our proposed architecture has a complexity of $\mathcal{O}(n)$ --or $\mathcal{O}(1)$ for a single point query-- and has no parallelization issues. We will show how this permits accurate and efficient collision handling in physically based simulations, more specifically, for cloth. In our experiments, we query up to $1$M points in $\sim300$ milliseconds. 
\end{abstract}

%%%%%%%%% BODY TEXT
\section{Introduction}

Physically based simulation has been a research topic for decades. It has numerous applications, ranging from realistic simulations for scientific purposes to the entertainment industry --films, videogames and VR/AR-- where accuracy is often relegated to a second plane in favor of looks and performance. Nonetheless, regardless of the application, simulations require accurate collision handling. With the development of faster GPUs and GPGPUs, researchers have leveraged the parallelization capabilities of these devices to speed up simulations. Nonetheless, collisions remain a major bottleneck.

Given two meshes representing different objects, a naive collision handling approach consists on testing each vertex-triangle pair between meshes. This becomes a very inefficient approach as the number of triangles grow, thus, simulation usually relies on acceleration structures, such as BVH, KD-Tree, AABB-tree, R-tree and other hierarchical spatial partitions \cite{bergen1997efficient, ericson2004real, palmer1995collision}. Hierarchical solutions significantly reduce computational time for collision queries. Nonetheless, they present some drawbacks. As aforementioned, researchers have long proposed moving simulations to GPUs and benefit from parallel computation. While acceleration structures work well in CPUs, due to GPUs architecture and the branching nature of these algorithms, the parallelization capabilities of GPUs cannot be fully exploited. Some works propose modified versions of BVH for ray-tracing that better adapts to GPUs and processes dense packets of rays to mitigate branching \cite{gunther2007realtime, hapala2011efficient}. We propose applying deep implicit surfaces as Signed Distance Functions (SDF) w.r.t. a watertight 3D mesh to handle collisions. This approach has a constant query time and does not suffer from branching, and thus, it effectively utilizes GPUs parallel computational power. It is important to note that the benefits of SDF for collision handling had already been studied~\cite{fuhrmann2003distance, macklin2020local}. These approaches compute discrete SDFs by voxelizing the space around a mesh and storing the signed distance in each cell. While these approaches are very efficient, they present a compromise between accuracy and memory footprint, as it is usual for voxelization.

Through this paper we will show how implicit surfaces learnt through neural networks --neural implicit surfaces from now on-- can be applied for efficient and accurate collision handling for physically based simulations. Our contributions w.r.t. previous methodologies are as follows:
\begin{itemize}
    \item \textbf{Efficient parallelization.} The methodology we present better exploits parallelization capabilities of current GPUs and GPGPUs. As opposed to acceleration structures, neural networks do not suffer from branching nor inefficient memory fetches.
    \item \textbf{Compact and continuous representation.} Neural SDF permit an accurate compact representation of continuous and differentiable surfaces. This is an important advantage against voxelized SDF, where accuracy comes at a cubic cost in terms of memory. Moreover, such discrete representations allows only for approximations of gradients --surface normals-- through finite differences where, again, its accuracy depend on the chosen voxel resolution, with its associated memory cost.
\end{itemize}

The rest of the paper is as follows. First, we review the related work on collision handling and deep implicit representations. Then, we describe the proposed methodology. Later, we define the experimental setup and present the obtained results. Finally, we close with the limitations found and set a possible future research line.

%------------------------------------------------------------------------
\section{Related work}

\textbf{Computer graphics.} Collision detection and solving is a crucial part on many computer graphics tasks. For this reason, this problem has been tackled for a long time. The first, most naive, solution is to apply brute force to the problem. Given two 3D objects represented by triangular meshes, we can compare each possible vertex-triangle pair. While this solution works well and is easy to implement, it has scalability issues, as its complexity grows with the square of the number of vertices and triangles to check. In order to reduce the cost of collision detection, acceleration structures have been proposed \cite{ericson2004real, palmer1995collision, bergen1997efficient}. The main idea behind these approaches is to partition the space hierarchically. Different works proposed different strategies --spheres-trees, BVH, KD-Tree, Oct-tree, AABB-trees, ...-- to achieve logarithmic search times. There had been efforts on building dynamically updated structures of this kind \cite{kopta2012fast} for animated scenes. Later, with the emergence of GPUs and GPGPUs, researchers proposed moving simulations into these devices to benefit from their parallel computation capabilities and speed up the algorithms. Nonetheless, collision detection remained a challenge. This is due to the architecture of GPUs --which we still find in modern GPUs-- not being able to parallelize branched algorithms. Authors proposed modified acceleration structures that better suit GPUs for ray-tracing \cite{gunther2007realtime, hapala2011efficient, afra2014stackless}. These solutions rely on the possibility of processing dense packets of rays in parallel. Rays of light close to each other will most likely intersect with the same bounding volumes, and thus, minimize the branching in this kind of algorithms. Whenever rays diverge, they are usually stored in a GPU stack memory, that in turn, presents issues of its own due to not being optimized for this task. Finally, we find Signed Distance Functions, computed as voxels in which the signed distance to a 3D object is stored in each cell \cite{fuhrmann2003distance, macklin2020local}. These solutions allow for efficient collision solving without branching. Their main drawback is a cubic memory footprint cost w.r.t. the chosen voxel resolution, and thus, a compromise between accuracy and resources. Thus, a general purpose parallelizable collision handling algorithm remains an open problem. We propose using neural implicit surfaces for collision solving. Neural networks do not suffer from branching and therefore, can fully exploit GPU parallel computation capabilities. Furthermore, they offer a compact representation for 3D shapes.

\textbf{Divergence and coalescence.} We have mentioned how GPU architectures are not suitable for acceleration structures. We can identify two main drawbacks regarding these solutions. First, divergence. GPU cores are grouped into blocks (32 for Nvidia and 64 for AMD). These blocks need to compute the exact same operations, then, whenever an algorithm requires branching (\textit{if-else} statements and similar), unless all threads follow the same \textit{path}, the code corresponding to each branch shall be executed sequentially. This is called divergence. Acceleration structures perform searches by branching, and thus, they suffer from divergence in GPU. Then, we have coalescence. Memory fetches performed per each block are done sequentially. Each I/O operation will gather a chunk of memory ($32$ or $128$ bytes, depending on the architecture) regardless of the required data size. Then, if a block of threads requests data placed in the same memory chunk, a single memory fetch can feed all threads. On the other hand, if data to be read is spread through memory, each thread will require individual memory fetches, that are performed sequentially, which has a big impact on performance (I/O operations are costly). Memory coalescing happens whenever consecutive threads work with consecutive data (single fetch). Acceleration structures will also require inefficient memory fetches, which will impact the performance. On the other hand, note how neural networks do not suffer from divergence and can benefit from coalescence.

\textbf{Deep implicit representations.} Recently, deep learning research has shown interest in implicit representations. Authors of \cite{sitzmann2020implicit} propose an architecture capable of accurately capturing detailed implicit functions in image domain, audio domain and 3D domain by using periodic activation functions only. Also, in \cite{tancik2020fourier}, authors show how using a Fourier feature mapping allows learning implicit functions with high fidelity and detail, also in image and 3D domain. Then, the work of \cite{park2019deepsdf} proposes learning Signed Distance Functions with neural networks. Moreover, they show it is possible to condition the input of the model to generate SDF for different 3D shapes. Following a similar approach, authors of \cite{corona2021smplicit} proposed an Unsigned Distance Function for open meshes to allow generation of different 3D clothes with a single model. These works propose, as an application, the possibility of reconstructing incomplete 3D data with their learnt SDF --partial point clouds or depth image estimations-- to regress full 3D shapes. Then, in \cite{santesteban2021self}, within garment domain as well, authors propose using a shallow MLP to learn an SDF of a human body with neutral shape in rest pose to implicitly solve predicted garment collisions by enforcing a constraint as a loss function during network training. To the best of our knowledge, we are the first to propose leveraging neural implicit surfaces --as SDF-- to explicitly solve collisions within physically based simulations.

%------------------------------------------------------------------------
\section{Methodology}

In this section we will describe the methodology used to learn SDFs and how to apply them to accurately handle collisions in a physically based simulation.

\subsection{Neural SDF}

The current literature already explores the possibility of learning implicit representations through deep learning \cite{sitzmann2020implicit, tancik2020fourier, park2019deepsdf}. In our case, we want to learn a MLP for each individual 3D object. We follow a similar strategy as \cite{park2019deepsdf}. We sample points near the object surface and also uniformly in the space around it. Then, we compute the signed distance for each of these points and learn a MLP to map point coordinates to its corresponding signed distance. That is:
\begin{equation}
    f: \mathbf{p}\in\mathbb{R}^3\rightarrow d\in\mathbb{R}
\end{equation}
Where $\mathbf{p}$ is the point in the 3D space and $d$ is the signed distance. Since collisions happen near surface-level, we do not need accurate predictions for points that are not close to the surface. We achieve this by balancing the ratio of sampled points near the surface w.r.t. uniform space sampling. We observed that a ratio of $80/20$ yields good results. Surface sampling margin is between $.5$cm and $1$cm, depending on the global size of the object. We try two different architectures for our neural network: a baseline MLP and a MLP with Fourier features\cite{tancik2020fourier}. For training our network, we choose a standard L1 loss. Note how a L2 loss prioritizes learning signed distances with larger error --larger gradient-- which will most likely be far away from the surface. An L1 loss has a constant gradient and treats equally all errors. Then, due to the balance proposed for the sampled points, the region near the surface is more accurately learnt.

\subsection{Collision solving}
\label{sec:collision_solving}

Once an accurate SDF has been learnt, collision detection and solving can be performed. Detection is trivial. We define the collision condition as $f(\mathbf{p}) < \epsilon$. We choose $\epsilon = 1$mm in our experiments. Then, to compute the collision point we leverage the differentiability of neural networks. First, we compute the surface normal as:
\begin{equation}
    \mathbf{N}_\mathbf{p} = \frac{\nabla f(\mathbf{p})}{ \|\nabla f(\mathbf{p})\|}
    \label{eq:surf_normal}
\end{equation}
Then, we compute the contact point as:
\begin{equation}
    \mathbf{x} = (\epsilon - f(\mathbf{p}))\mathbf{N}_\mathbf{p}
    \label{eq:contact_point}
\end{equation}
Note that the direction of the gradient might need to be reversed depending on the convention used for interior/exterior. In our experiments, we define the interior of the mesh as negative, such that the gradient naturally points outwards. Finally, thanks to the back-propagation technique, the gradient can be efficiently computed. We apply a naive collision solving strategy where we modify the collided vertex locations to the contact point $\mathbf{x}$. This simple setup is enough to prove the capacity of neural implicit surfaces to accurately determine the contact point. We consider advanced collision handling techniques to be outside of the scope. Note that, if the contact point can be accurately computed, this methodology is compatible with more complex collision solving algorithms\cite{baraff1998large, macklin2020local}.

\section{Experiments}

In this section we will explain the experimental setup and obtained results. We test our methodology for cloth simulation.

\subsection{Cloth simulator}

We implement a simple cloth simulator in TensorFlow. We design our simulator for rectangular pieces of cloth with quad faces. We choose an explicit Verlet integrator. Note that we limit the scope of our work to collisions only. Thus, we aim to assess only the performance and accuracy of the proposed collision handling system. We leverage the automatic differentiation tool of TensorFlow to compute the gradient of $f$, and ultimately, the collision point as described in Eq. \ref{eq:contact_point}. Then, we perform analogous simulations using: a) an acceleration structure and b) a voxelized SDF. An acceleration structure does not require computing the gradient, as a surface normal can be analytically computed. Then, for voxelized SDF there are two options: pre-computing gradients and storing them in each cell --increasing memory footprint by a factor of $4$-- or computing it on the fly with finite differences. Note how both approaches assume whether constant gradient or constant distance within each cell, which shall yield an error inversely proportional to the chosen voxel resolution.

\subsection{Results}

\begin{figure}
    \centering
    \includegraphics[width=\columnwidth]{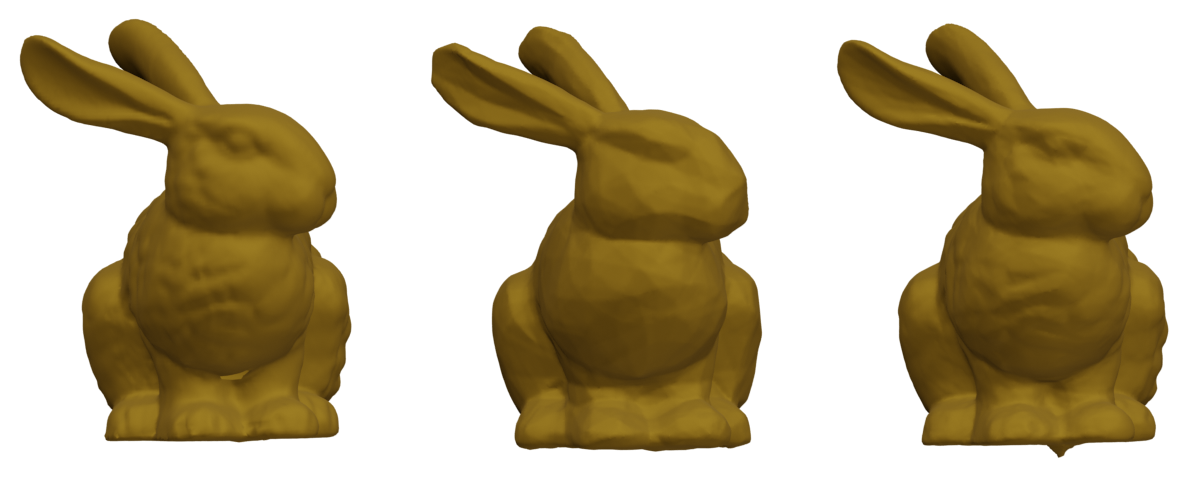}
    \caption{Stanford bunny reconstruction from neural implicit functions using marching cubes algorithm. From left to right: ground truth, baseline MLP and Fourier features \cite{tancik2020fourier}.}
    \label{fig:bunny_reconstruction}
\end{figure}

\begin{figure*}
    \centering
    \includegraphics[width=\textwidth]{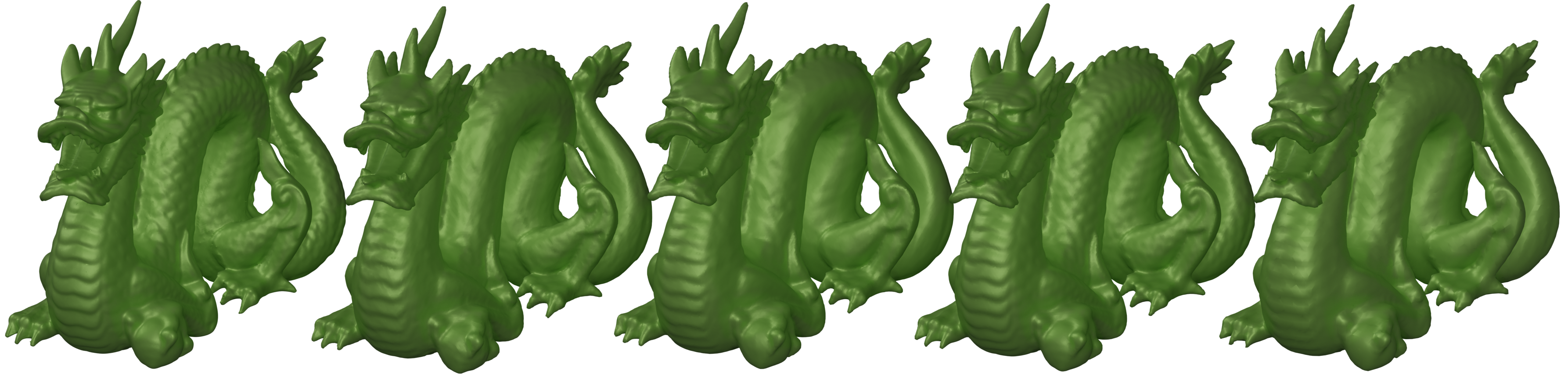}
    \caption{Stanford dragon reconstruction from implicit functions using marching cubes algorithm. From left to right: ground truth, baseline MLP, Fourier features \cite{tancik2020fourier}, voxelized SDF with 256 voxels and voxelized SDF with 128 voxels.}
    \label{fig:dragon_reconstruction}
\end{figure*}

\begin{figure*}[ht!]
    \centering
    \includegraphics[width=\textwidth]{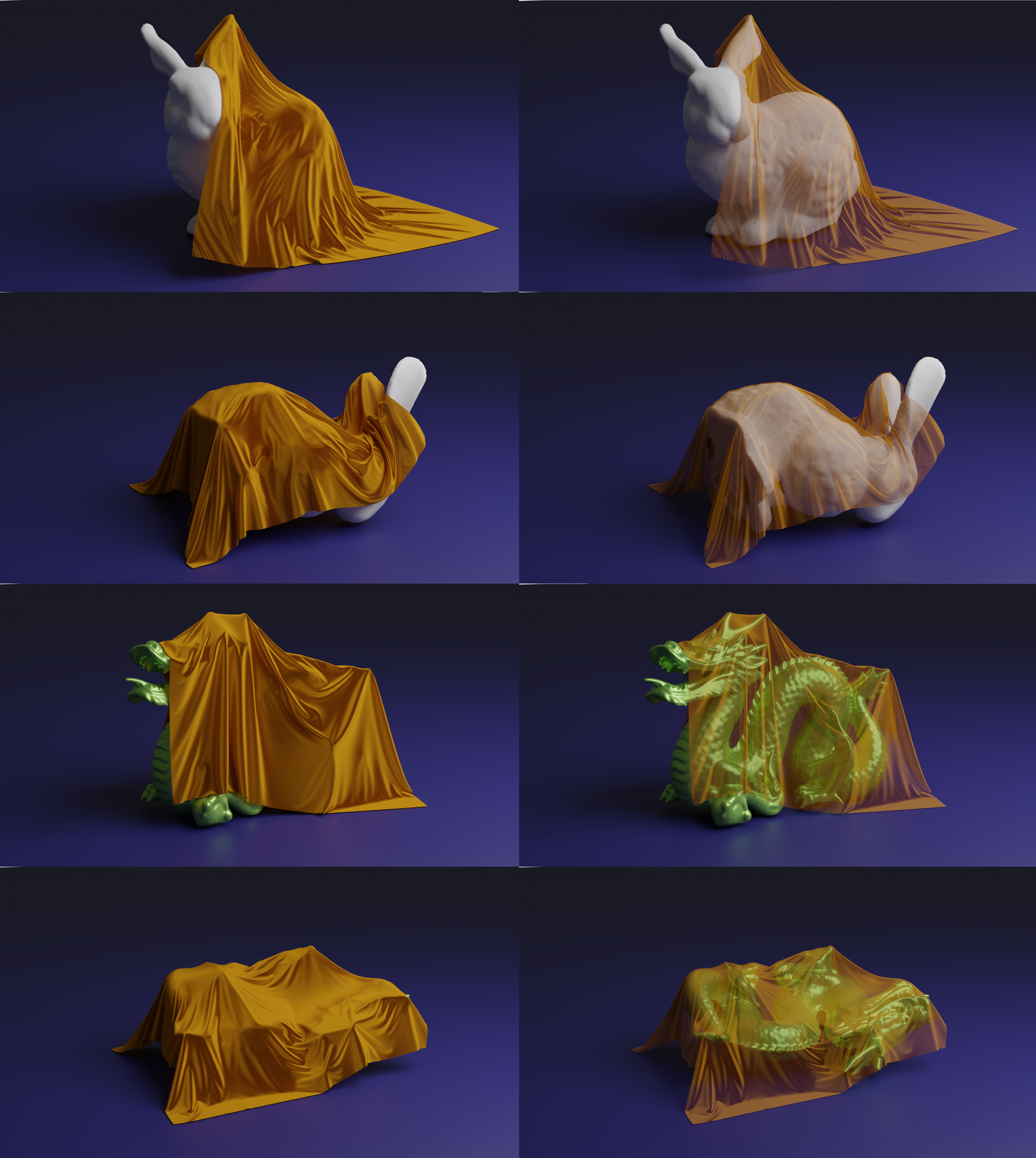}
    \caption{Qualitative results obtained with our cloth simulator by implementing collisions as explained in Sec. \ref{sec:collision_solving}. While we use the learnt SDF to solve collisions, we use the original 3D mesh for rendering the results. This shows how well aligned is the SDF w.r.t. original mesh. For each sample, we render cloth with and without opacity to better illustrate the accuracy of the results.}
    \label{fig:qualitative}
\end{figure*}

First, we analyze the accuracy of the learnt implicit representations by visualizing the mesh reconstruction. To this end, we use the marching cubes algorithm. We compare the representation capabilities of two different approaches: first, a simple MLP as a baseline and secondly, a MLP with Fourier features as input \cite{tancik2020fourier}. Both models contain $4$ fully connected layers with $256$ hidden units each. Fig. \ref{fig:bunny_reconstruction} shows the obtained results. As observed, a simple MLP can learn the overall shape of the object, but cannot capture the details. Then, mapping input features to Fourier features allows learning even smallest details. Then, we perform a similar experiment with the stanford dragon model. This model contains a much larger number of triangles and level of details. We observed $4$ layers are not enough to capture its geometry, thus, in this experiments, we use MLP with $8$ layers. Fig. \ref{fig:dragon_reconstruction} shows the results obtained. We observe that our simple baseline can capture small details for this experiment. We observe no difference to the Fourier features approach. Nonetheless, neither of both is able to capture the finest details, the scales in the body and the wrinkles in the head. We will show how it is still possible to use them for accurate collision solving. Finally, we see how a voxelization with a resolution of $256$ voxels outperforms both of our neural networks, while $128$ voxels show a slight loss of details, as can be seen in the face and horns of the dragon. The original model has a size of $34.8$MB, while both MLP have a size of $2.01$MB. Finally, the voxelized SDFs have a size of $64$MB and $8$MB respectively. The average reconstruction error for all of them is below $0.1$mm. Thus, since we defined $\epsilon = 1mm$, we consider them accurate enough to be used in collision handling.

We perform the cloth simulation as explained previously with a piece of cloth with $N = 40000$ vertices. We present the results in Fig. \ref{fig:qualitative}. Note that the shown object corresponds to the ground truth 3D mesh, while collision is handled by the implicit surface learnt through neural networks. For each sample, we render the piece of cloth with and without opacity to enhance the visibility. As can be observed, the methodology proposed in this paper allows for accurate collision handling. Finally, as it is common with standard SDFs and acceleration structures, it is also possible to rotate, scale or move the neural implicit surfaces by transforming cloth vertices into the local coordinate system of our SDF. That is, given a matrix $\mathbf{T}\in\mathbb{R}^{4\times4}$ describing a rotation and translation for our 3D object, we substitute $\mathbf{p}$ by $\mathbf{p}' = \mathbf{T}^{-1}\mathbf{p}$ in Eq. \ref{eq:surf_normal} and Eq. \ref{eq:contact_point}. Finally, $\mathbf{N}_\mathbf{p}$ must be rotated accordingly with $\mathbf{T}_{:3,:3}\in\mathbb{R}^{3\times3}$. Note that this is equivalent to the transformation $\mathbf{T}$ without translation, since $\mathbf{N}_\mathbf{p}$ is a normal vector.

\subsection{Performance}

\begin{table*}[]
\centering
\begin{tabular}{c|c|c|c|c|c|c|}
\cline{2-7}
 & \multicolumn{2}{c|}{KD-Tree} & \multicolumn{2}{c|}{Voxel SDF ($N=256$)} & \multicolumn{2}{c|}{Neural SDF} \\ \hline
\multicolumn{1}{|c|}{N. queries} & Bunny & Dragon & Bunny & Dragon & Bunny & Dragon \\ \hline
\multicolumn{1}{|c|}{10K} & 11.7ms (675ms) & 51.9ms (748ms) & $\sim0.1$ms & $\sim0.1$ms & 0.73ms & 1.00ms \\ \hline
\multicolumn{1}{|c|}{40K} & 45.3ms (4.2s) & 209.9ms (4.4s) & $\sim0.1$ms & $\sim0.1$ms & 4.35ms & 0.95ms \\ \hline
\multicolumn{1}{|c|}{1M} & 1415ms (173s) & 9900ms (175s) & $\sim0.1$ms & $\sim0.1$ms & 350.0ms & 255.0ms \\ \hline
\end{tabular}
\caption{Computational times for collision queries using different methodologies: a) acceleration structure (with and without parallelization), b) voxelized SDF and c) neural SDF. We check against two different 3D objects, the stanford bunny and dragon. Note that the dragon has over $10$ times the number of triangles of the bunny. This has an impact in acceleration structures and neural implicit surfaces. Voxelization has been computed with a resolution of $N = 256$ and it is agnostic to the number of triangles.}
\label{tab:performance}
\end{table*}

We analyze the performance of the three main approaches discussed in this paper for collision handling. That is: acceleration structures, voxelized SDFs and neural SDFs. We replicate the experimental setup for collision handling in cloth simulation. All of the methodologies allow computing a signed distance and a normal vector, and thus are compatible with Eq. \ref{eq:contact_point}. We found no observable qualitative differences between the simulation outcomes. Tab. \ref{tab:performance} reports the performance of the different methodologies tested with two different 3D objects: the stanford bunny and dragon. We consider only the time required to obtain the contact point (Eq. \ref{eq:contact_point}). For KD-Tree, we use the \textit{cKDTree} implementation available for Python. It is efficiently coded in C++ and allows for parallel queries in multi-thread CPUs. We test this solution on an Intel Xeaon Platinum 8269CL CPU with $48$ cores. We report numbers for parallel and serialized queries respectively. As observed, this algorithm hugely benefits from the multi-core nature of modern CPUs. As previously discussed, while acceleration structures for GPU exist, their focus is on ray-tracing, where packets of rays can be processed in parallel with minimal divergence. Additionally, we observe that more complex 3D meshes --larger number of triangles-- results on longer querying times. Next, we report the results for classical voxelized SDF. In this case, obtaining the signed distance of a given point can be computed with a single reading operation, since SDF is encoded like a hash table. Also, the memory cost of this solution depends only on the chosen resolution. In practice, we find similar running times regardless of the number of queries and the 3D object. The reported numbers are due to the overhead of calling TensorFlow functions from Python, thus, querying times are too small to measure accurately. This is the expected results for this solution. Then, the only drawback of voxelized SDF on run-time is their memory footprint. Also, while a resolution of $N = 256$ is enough for the tested objects, this methodology would present scalability issues when applied to whole scenes --like a videogame map or a large rendering scenario-- and larger resolutions are required. Finally, we present the results obtained with the proposed neural networks. We run the experiments in a GPU Tesla T4. In each network, we have a first \textit{layer} where features are transformed into Fourier features \cite{tancik2020fourier}. Then, the bunny SDF has $4$ fully connected layers and the dragon has $8$. All of the layers have $256$ hidden units and ReLu activation function. We observe the computation times are lower than parallel searching through acceleration structures for all cases. On the other hand, it is slower than a voxelized solution (as expected). We can also observe how, as it is natural, the larger the number of queries, the longer it requires. Nonetheless, we found an interesting observation. For larger number of queries, the dragon model --with more layers-- takes less time than the bunny model. This is due to optimization techniques used by TensorFlow during back-propagation. Whenever a ReLu activation yields $0$, the gradient is not computed. Thus, for deeper models, we will have more gradients dropped, and overall, lower back-propagation times. We observed that for a large number of queries, the back-propagation is significantly more costly than the forward pass. We explored the idea of predicting the normal of the closest surface along with the signed distance, but this space shows numerous discontinuities (imagine the space in between two walls, discontinuities would appear in the middle points), and thus, we discarded this approach. In conclusion, we showed how neural SDF can improve computational times w.r.t. acceleration structures. On the other hand, voxelized SDF is unlikely to be outperformed in run-time. Nonetheless, we also showed how neural SDF presents a much more compact representation in terms of memory (bunny SDF: $0.76$MB, dragon SDF: $2.01$MB, voxelization with $N = 256$: $64$MB) which might have an impact in the scalability of these solutions.

It is also important to take into account the time it takes for the construction of each model. For acceleration structures, the cost of building the tree is $\mathcal{O}(n)$, where $n$ is the number of vertices or triangles on the mesh. In this sense, it is the fastest approach. On the other hand, note how voxelized SDF require to sample $\mathcal{O}(n^3)$ points, where $n$ is the chosen resolution. For $n = 256$, this is almost $17$M of point queries to construct the SDF. For neural SDF, since we want them to be accurate around the mesh, we do not need to uniformly sample in a three dimensional space. In practice, most 3D objects are represented by two dimensional surfaces, and thus, we can consider the amount of required sample points to grow as $\mathcal{O}(n^2)$ w.r.t. mesh size. In the work of \cite{park2019deepsdf}, they propose sampling around $250$K points per shape. We find this to be insufficient for the level of accuracy required for simulations. Then, to ensure accuracy, we sampled around $1$M points for the bunny and $5$M points for the dragon. Finally, due to the simplicity of the architecture, we find training times of a few tens of minutes. Nonetheless, overall, we devoted more time into obtaining neural SDFs than other representations.

\section{Limitations and Conclusions}

We proposed an overlooked application for neural implicit surfaces formulated as Signed Distance Functions. We demonstrated how it is possible to use them for efficient and accurate collision handling in physically based simulations. We compared this approach with standard computer graphics techniques: acceleration structures and voxelized SDFs. We concluded that neural SDF falls in between both of these approaches in terms of efficiency. Neural SDF can achieve faster running times than acceleration structures, but cannot beat voxelized solutions. On the other hand, we saw how neural networks offer a much more compact representation for SDF. Neural SDF presents a limitation we also find in standard voxelized SDF. They are not suitable for self-collisions, which is an important part of certain simulations (particle systems). Finally, neural networks have the potential capability of representing multiple SDF with a single model. We tried learning 4D objects (3D objects in time) using neural networks. Unfortunately we did not achieve the necessary level of accuracy to perform cloth simulations. Nonetheless, we believe it is an interesting future research line that will set a clear advantage of neural SDF over classical voxelized SDF.

{\small
\bibliographystyle{ieee_fullname}
\bibliography{egbib}
}

\end{document}